# Introduction Of Quantum Entanglement Measure Based On The Expectation Values Of Pauli Operators

**Mahmood Zeheiry (zeheiry.m@gmail.com)**

Department of Physics, Faculty of Science, Shahid Chamran University of Ahvaz, Ahvaz, Iran.

**Abstract**

In this paper, firstly considering that in separable states, the measurement of one particle has no effect on the measurement of the second particle, we show that Alice and Bob can find directions in which the results of their measurements on the spin of the particle are always maximized. In other words, the state of the particle is an eigenstate for the operator that is applied in that direction, so the sum of the spins of two particles can have a maximum value. We will argue that in entangled states, due to the effect of particle measurement results on each other, Alice and Bob cannot find the desired operators. Therefore, in such measurements, the total spin of the particles will always be less than the mentioned maximum. But we ask them to try and measure in directions that will get the most value. Because this value is maximum for separable states and minimum for fully entangled states, and for the rest of the states, it will be proportional to the degree of entanglement between the two maximum and minimum values, we set this parameter as We are calling it the "separability index". Then, based on this index, the measure of entanglement was introduced and extended to states with higher dimensions. In the end, examples of di-qubit states di-qutrit states, and di-qudit states were investigated and the efficiency of the measure was confirmed by the results of the examples. Considering that in this measure, the values of entanglement are calculated based on the expectation values and we can measure the expectation values in the experiment, we hope to be one step closer to the testability of the entanglement value.

**Introduction:**

Entanglement is one of the most suitable resources for quantum information processing and also it has many applications in quantum information and quantum field theories [1]. For this reason, knowing the entangled states and determining their entanglement value is of great importance. Considering the importance of entanglement in quantum mechanics; various measures were introduced to determine the degree of entanglement of states. Among these measures, we can mention the geometric measure that was first introduced by Abner Shimoni in 1995 based on the minimum distance between the desired state and separable states [2]. Then, in 2015, the spin expectation value method was used as a method for calculating the geometric measure for qubit states, which considers the multi-qubit state as a two-component state and the entanglement value of one component. Calculates with the rest of the system [3].

In this paper, we intend to introduce a measure for qubit states without using relations and concepts of geometric measure and completely independently, by using the effect of entanglement on the expectation values of Pauli operators. Then we generalize the problem for qudit states as well. The advantage of this measure is that not only it is based on the concepts of entanglement, but it also calculates the value of entanglement for the entire system and it is not like this that calculates the amount of entanglement of a component with other components of the system.

## 1. The qubit states are eigenstates of Pauli operators

Consider the state $|\psi\rangle = \frac{1}{\sqrt{2}}(|0\rangle + |1\rangle)$, which $|0\rangle$ and $|1\rangle$ are eigenstates for the $\sigma_z$ operator with 1 and -1 eigenvalues. The expectation value of this operator on this state is 0, but we know that it is an eigenstate of the $\sigma_x$ operator, that maximizes the expectation value of this operator. So, we can rotate the Stern-Gerlach device in the $x$ direction and after measuring, get the expectation value of 1. Generally, for every arbitrary qubit state $|\chi\rangle = \alpha|0\rangle + \beta|1\rangle$ direction like n can be found, where the state $|\chi\rangle$ is the eigenstate of the operator $\sigma_n$ with the eigenvalue of 1. The expected value of the operator $\sigma_n$ in this state will be 1. Finding this operator and finding this direction is an easy task for single-qubit states.

## 2. What about two-qubit states?

Now we examine the problem for two-qubit states. First, we consider a separable two-qubit state $|\psi\rangle = |A\rangle|B\rangle$. Suppose: qubit A is in the possession of Alice and qubit B is in the possession of Bob. Alice can always find an $\sigma_1$ operator whose $|A\rangle$ is an eigenstate with an eigenvalue equal to 1 and also Bob can find an $\sigma_2$ operator for $|B\rangle$. To calculate the sum of expectation values of operators $\sigma_1$ and $\sigma_2$ in $|\psi\rangle$; we have:

$$\langle\psi|\sigma_1|\psi\rangle + \langle\psi|\sigma_2|\psi\rangle = \langle AB|\sigma_1|AB\rangle + \langle AB|\sigma_2|AB\rangle = 2 \qquad (1)$$

Now let us consider the fully entangled state $|\phi\rangle = \frac{1}{\sqrt{2}}(|01\rangle + |10\rangle)$, For this state, according to the concept of entanglement, Alice cannot find an operator whose state $|\phi\rangle$ is its eigenstate, because in this case, after measuring, Alice always assigns a value (the corresponding eigenvalue) to he gets it and basically, such a thing is against the definition and concept of entanglement, which says that the measurement results of Alice and Bob are dependent on each other. Bob will have a similar situation.

Therefore, for entangled states, the sum of expectation values for the whole system becomes less than the value of the relation (1). We divide the sum of the maximum expectation values that Alice and Bob obtain by rotating the measurement direction among the components of the system and call it the "separability index" of the system, which is generally for systems N-Partial is defined as follows:

$$\gamma = \frac{\max\langle\sigma_1\rangle + \max\langle\sigma_2\rangle + ... \max\langle\sigma_N\rangle}{N} \qquad (2)$$

With this definition, $\gamma$ will always be equal to 1 for separable 2-qubit states.

The question may be asked why, in the case of deducting the relationship (2), the maximum expected values were used. In the answer, it should be said that for separable states, Bob and Alice have the freedom to choose the Pauli operators to obtain the expectation values of each qubit between -1 and 1, that is, for the sum of the two-qubit expectation values, It varies in $[-2,2]$ range. But for entangled states, this freedom of action is lost, and depending on the amount of entanglement, the sum of expectation values will be a number in the $[-2,2]$ range, and the higher the entanglement of the state, the closer this number is to the center of the interval (zero here), so that for the completely entangled state, this value is equal to the average of the eigenvalues, that is, it becomes zero. Therefore, to distinguish between separable and entangled states, we ask Alice and Bob to try to find a direction for each of them that will measure the highest possible expectation value of their qubit.

For the separability index to fulfill our purpose well, its value should be zero for fully entangled qubit states. For this purpose, we examine the relation (1) for the fully entangled state $|\phi\rangle = \frac{1}{\sqrt{2}}(|01\rangle + |10\rangle)$ in the matrix representation:

First, we obtain the $\sigma_1$ and $\sigma_2$ matrices [4].

According to $\sigma_i = \sigma_x \sin\theta_i \cos\varphi_i + \sigma_y \sin\theta_i \sin\varphi_i + \sigma_z \cos\theta_i$ where $\theta_i$ is the polar angle that the $\sigma_i$ operator makes with the $z$ axis in the Blach sphere and $\varphi_i$ is the directional angle that the $\sigma_i$ operator makes with the $x$ axis in the Blach sphere [4]:

$$\sigma_1 = \sigma_1 \otimes I = \begin{pmatrix} \cos\theta_1 & 0 & e^{-i\varphi_1}\sin\theta_1 & 0 \\ 0 & \cos\theta_1 & 0 & e^{-i\varphi_1}\sin\theta_1 \\ e^{i\varphi_1}\sin\theta_1 & 0 & -\cos\theta_1 & 0 \\ 0 & e^{i\varphi_1}\sin\theta_1 & 0 & -\cos\theta_1 \end{pmatrix} \qquad (3)$$

$$\sigma_2 = I \otimes \sigma_2 = \begin{pmatrix} \cos\theta_2 & e^{-i\varphi_2}\sin\theta_2 & 0 & 0 \\ e^{i\varphi_2}\sin\theta_2 & -\cos\theta_2 & 0 & 0 \\ 0 & 0 & \cos\theta_2 & e^{-i\varphi_2}\sin\theta_2 \\ 0 & 0 & e^{i\varphi_2}\sin\theta_2 & -\cos\theta_2 \end{pmatrix} \qquad (4)$$

$$\langle\phi|\sigma_1|\phi\rangle = \begin{pmatrix} 0 & \frac{1}{\sqrt{2}} & \frac{1}{\sqrt{2}} & 0 \end{pmatrix} \begin{pmatrix} \cos\theta_1 & 0 & e^{-i\varphi_1}\sin\theta_1 & 0 \\ 0 & \cos\theta_1 & 0 & e^{-i\varphi_1}\sin\theta_1 \\ e^{i\varphi_1}\sin\theta_1 & 0 & -\cos\theta_1 & 0 \\ 0 & e^{i\varphi_1}\sin\theta_1 & 0 & -\cos\theta_1 \end{pmatrix} \begin{pmatrix} 0 \\ \frac{1}{\sqrt{2}} \\ \frac{1}{\sqrt{2}} \\ 0 \end{pmatrix} = 0 \quad (5\text{-a})$$

$$\langle\phi|\sigma_2|\phi\rangle = \begin{pmatrix} 0 & \frac{1}{\sqrt{2}} & \frac{1}{\sqrt{2}} & 0 \end{pmatrix} \begin{pmatrix} \cos\theta_2 & e^{-i\varphi_2}\sin\theta_2 & 0 & 0 \\ e^{i\varphi_2}\sin\theta_2 & -\cos\theta_2 & 0 & 0 \\ 0 & 0 & \cos\theta_2 & e^{-i\varphi_2}\sin\theta_2 \\ 0 & 0 & e^{i\varphi_2}\sin\theta_2 & -\cos\theta_2 \end{pmatrix} \begin{pmatrix} 0 \\ \frac{1}{\sqrt{2}} \\ \frac{1}{\sqrt{2}} \\ 0 \end{pmatrix} = 0 \quad (5\text{-b})$$

Therefore, the separability index ($\gamma$) becomes zero for the fully entangled state.

For the state $|\psi\rangle = \frac{1}{2}(|00\rangle + |01\rangle + |10\rangle + |11\rangle)$, the expectation values will be as follows:

$$\langle\sigma_1\rangle = \cos\varphi_1 \sin\theta_1 \quad \Rightarrow \max\langle\sigma_1\rangle = 1 \quad (6)$$

$$\langle\sigma_2\rangle = \cos\varphi_2 \sin\theta_2 \quad \Rightarrow \max\langle\sigma_2\rangle = 1 \quad (7)$$

So separability index is equal to 1 and the mentioned state is a separable state.

## 3. definition of entanglement measure

Now, we can define the entanglement measure by using the separability index as follows:

$$E = 1 - \gamma \quad (8)$$

According to this relation, the value of the measure for the separable and fully entangled states introduced in section 2 will be zero and 1, respectively.

This measure has all the necessary characteristics, including uniformity of entanglement under LOCC transformations [5], maximum value for Bell states, and zero value for separable states.

*Example 1:* We calculate the entanglement of the state $|\psi\rangle = \frac{1}{2}|00\rangle + \frac{\sqrt{3}}{2}|11\rangle$:

$$\langle\psi|\sigma_1|\psi\rangle = \frac{1}{4}\begin{pmatrix}1 & 0 & 0 & \sqrt{3}\end{pmatrix}\begin{pmatrix} \cos\theta_1 & 0 & e^{-i\varphi_1}\sin\theta_1 & 0 \\ 0 & \cos\theta_1 & 0 & e^{-i\varphi_1}\sin\theta_1 \\ e^{i\varphi_1}\sin\theta_1 & 0 & -\cos\theta_1 & 0 \\ 0 & e^{i\varphi_1}\sin\theta_1 & 0 & -\cos\theta_1 \end{pmatrix}\begin{pmatrix}1\\0\\0\\\sqrt{3}\end{pmatrix} = -\frac{1}{2}\cos\theta_1$$

$$\langle\psi|\sigma_2|\psi\rangle = \frac{1}{4}\begin{pmatrix}1 & 0 & 0 & \sqrt{3}\end{pmatrix}\begin{pmatrix} \cos\theta_2 & e^{-i\varphi_2}\sin\theta_2 & 0 & 0 \\ e^{i\varphi_2}\sin\theta_2 & -\cos\theta_2 & 0 & 0 \\ 0 & 0 & \cos\theta_2 & e^{-i\varphi_2}\sin\theta_2 \\ 0 & 0 & e^{i\varphi_2}\sin\theta_2 & -\cos\theta_2 \end{pmatrix}\begin{pmatrix}1\\0\\0\\\sqrt{3}\end{pmatrix} = -\frac{1}{2}\cos\theta_2$$

$$\gamma = \frac{\max\langle\sigma_1\rangle + \max\langle\sigma_2\rangle}{2} = \frac{1}{2} \quad \Rightarrow E = 1 - \frac{1}{2} = \frac{1}{2}$$

With this method, the value of $\frac{2}{5}$ is obtained for $|\chi\rangle = \frac{1}{\sqrt{5}}|00\rangle + \frac{2}{\sqrt{5}}|11\rangle$.

Now, for comparison, we calculate the entanglement of these states using the concurrence measure.

We must write the state in the form of the basis vectors of the two-component system as $|\psi\rangle = \sum_{i,j=0}^{1} a_{ij}|i,j\rangle$. Then concurrence measure is calculated as follows:[6,7]

$$C = 2|a_{00}a_{11} - a_{01}a_{10}| \Rightarrow C(\psi) = 2\left|\frac{1}{2} \times \frac{\sqrt{3}}{2} - 0\right| = \frac{\sqrt{3}}{2}$$

And concurrence measure for $|\chi\rangle$ is obtained:

$$C(\chi) = 2\left|\frac{1}{\sqrt{5}} \times \frac{2}{\sqrt{5}} - 0\right| = \frac{4}{5}$$

In the next example, we can see that, as mentioned in the introduction, unlike the method proposed in the article [3], which calculated the entanglement for only two components of the desired state, this measure, in the entanglement. It calculates in general and for all state components.

***Example 2:*** Entanglement computation of 3-qubit states:

At first, we obtain $\sigma_1$, $\sigma_2$, and $\sigma_3$ matrices.

$$\sigma_1 = \sigma_1 \otimes I \otimes I = \begin{pmatrix} \cos\theta_1 & 0 & 0 & 0 & e^{-i\varphi_1}\sin\theta_1 & 0 & 0 & 0 \\ 0 & \cos\theta_1 & 0 & 0 & 0 & e^{-i\varphi_1}\sin\theta_1 & 0 & 0 \\ 0 & 0 & \cos\theta_1 & 0 & 0 & 0 & e^{-i\varphi_1}\sin\theta_1 & 0 \\ 0 & 0 & 0 & \cos\theta_1 & 0 & 0 & 0 & e^{-i\varphi_1}\sin\theta_1 \\ e^{i\varphi_1}\sin\theta_1 & 0 & 0 & 0 & -\cos\theta_1 & 0 & 0 & 0 \\ 0 & e^{i\varphi_1}\sin\theta_1 & 0 & 0 & 0 & -\cos\theta_1 & 0 & 0 \\ 0 & 0 & e^{i\varphi_1}\sin\theta_1 & 0 & 0 & 0 & -\cos\theta_1 & 0 \\ 0 & 0 & 0 & e^{i\varphi_1}\sin\theta_1 & 0 & 0 & 0 & -\cos\theta_1 \end{pmatrix}$$

$$\sigma_2 = I \otimes \sigma_2 \otimes I = \begin{pmatrix} \cos\theta_2 & 0 & e^{-i\varphi_2}\sin\theta_2 & 0 & 0 & 0 & 0 & 0 \\ 0 & \cos\theta_2 & 0 & e^{-i\varphi_2}\sin\theta_2 & 0 & 0 & 0 & 0 \\ e^{i\varphi_2}\sin\theta_2 & 0 & -\cos\theta_2 & 0 & 0 & 0 & 0 & 0 \\ 0 & e^{i\varphi_2}\sin\theta_2 & 0 & -\cos\theta_2 & 0 & 0 & 0 & 0 \\ 0 & 0 & 0 & 0 & \cos\theta_2 & 0 & e^{-i\varphi_2}\sin\theta_2 & 0 \\ 0 & 0 & 0 & 0 & 0 & \cos\theta_2 & 0 & e^{-i\varphi_2}\sin\theta_2 \\ 0 & 0 & 0 & 0 & e^{i\varphi_2}\sin\theta_2 & 0 & -\cos\theta_2 & 0 \\ 0 & 0 & 0 & 0 & 0 & e^{i\varphi_2}\sin\theta_2 & 0 & -\cos\theta_2 \end{pmatrix}$$

$$\sigma_3 = I \otimes I \otimes \sigma_3 = \begin{pmatrix} \cos\theta_3 & e^{-i\varphi_3}\sin\theta_3 & 0 & 0 & 0 & 0 & 0 & 0 \\ e^{i\varphi_3}\sin\theta_3 & -\cos\theta_3 & 0 & 0 & 0 & 0 & 0 & 0 \\ 0 & 0 & \cos\theta_3 & e^{-i\varphi_3}\sin\theta_3 & 0 & 0 & 0 & 0 \\ 0 & 0 & e^{i\varphi_3}\sin\theta_3 & -\cos\theta_3 & 0 & 0 & 0 & 0 \\ 0 & 0 & 0 & 0 & \cos\theta_3 & e^{-i\varphi_3}\sin\theta_3 & 0 & 0 \\ 0 & 0 & 0 & 0 & e^{i\varphi_3}\sin\theta_3 & -\cos\theta_3 & 0 & 0 \\ 0 & 0 & 0 & 0 & 0 & 0 & \cos\theta_3 & e^{-i\varphi_3}\sin\theta_3 \\ 0 & 0 & 0 & 0 & 0 & 0 & e^{i\varphi_3}\sin\theta_3 & -\cos\theta_3 \end{pmatrix}$$

If our state is like this; $|\phi\rangle = \frac{1}{\sqrt{5}}|011\rangle + \frac{2}{\sqrt{5}}|100\rangle$:

$$\langle\phi|\sigma_1|\phi\rangle = -\frac{3}{5}\cos\theta_1 \quad , \quad \langle\phi|\sigma_2|\phi\rangle = \frac{3}{5}\cos\theta_2 \quad , \quad \langle\phi|\sigma_3|\phi\rangle = \frac{3}{5}\cos\theta_3$$

$$\gamma = \frac{\max\langle\sigma_1\rangle + \max\langle\sigma_2\rangle + \max\langle\sigma_3\rangle}{3} = \frac{\frac{9}{5}}{3} = \frac{3}{5} \quad \Rightarrow E = 1 - \frac{3}{5} = \frac{2}{5}$$

And for $|W\rangle = \frac{1}{\sqrt{3}}(|001\rangle + |010\rangle + |100\rangle)$:

$$\langle\phi|\sigma_1|\phi\rangle = \frac{1}{3}\cos\theta_1 \quad , \quad \langle\phi|\sigma_2|\phi\rangle = \frac{1}{3}\cos\theta_2 \quad , \quad \langle\phi|\sigma_3|\phi\rangle = \frac{1}{3}\cos\theta_3$$

$$\gamma = \frac{\max\langle\sigma_1\rangle + \max\langle\sigma_2\rangle + \max\langle\sigma_3\rangle}{3} = \frac{1}{3} \quad \Rightarrow E = 1 - \frac{1}{3} = \frac{2}{3}$$

To calculate the entanglement of these two three-qubit states by the concurrence measure, we should use the generalized form of the concurrence measure. Concurrence for pure three-component states is defined as follows [8]:

$$C_M(\rho) = \sqrt{2(1 - Tr(\rho_M^2))}$$

Where $\rho_M$ is the reduced density matrix of both arbitrary components. The density matrix of state $|\phi\rangle$ is as follows:

$$\rho_\phi = \frac{1}{5}|011\rangle\langle 011| + \frac{2}{5}|011\rangle\langle 100| + \frac{2}{5}|100\rangle\langle 011| + \frac{4}{5}|100\rangle\langle 100|$$

$$\langle 0|\rho_\phi|0\rangle = \frac{1}{5}\langle 0|011\rangle\langle 011|0\rangle + \frac{2}{5}\langle 0|011\rangle\langle 100|0\rangle + \frac{2}{5}\langle 0|100\rangle\langle 011|0\rangle + \frac{4}{5}\langle 0|100\rangle\langle 100|0\rangle = \frac{4}{5}|10\rangle\langle 10|$$

$$\langle 1|\rho_\phi|1\rangle = \frac{1}{5}\langle 1|011\rangle\langle 011|1\rangle + \frac{2}{5}\langle 1|011\rangle\langle 100|1\rangle + \frac{2}{5}\langle 1|100\rangle\langle 011|1\rangle + \frac{4}{5}\langle 1|100\rangle\langle 100|1\rangle = \frac{1}{5}|01\rangle\langle 01|$$

$$\Longrightarrow \rho_{AB} = \frac{4}{5}|10\rangle\langle 10| + \frac{1}{5}|01\rangle\langle 01| = \begin{pmatrix} 0 & 0 & 0 & 0 \\ 0 & \frac{1}{5} & 0 & 0 \\ 0 & 0 & \frac{4}{5} & 0 \\ 0 & 0 & 0 & 0 \end{pmatrix}$$

Similarly, for $\rho_{AC}$ and $\rho_{BC}$ we get:

$$\rho_{AC} = \frac{4}{5}|10\rangle\langle 10| + \frac{1}{5}|01\rangle\langle 01| = \begin{pmatrix} 0 & 0 & 0 & 0 \\ 0 & \frac{1}{5} & 0 & 0 \\ 0 & 0 & \frac{4}{5} & 0 \\ 0 & 0 & 0 & 0 \end{pmatrix}$$

$$\rho_{BC} = \frac{4}{5}|00\rangle\langle 00| + \frac{1}{5}|11\rangle\langle 11| = \begin{pmatrix} \frac{4}{5} & 0 & 0 & 0 \\ 0 & 0 & 0 & 0 \\ 0 & 0 & 0 & 0 \\ 0 & 0 & 0 & \frac{1}{5} \end{pmatrix}$$

$Tr(\rho_M^2)$ is $\frac{17}{25}$ for all three matrices. Then:

$$C_M(\rho) = \sqrt{2(1 - Tr(\rho_M^2))} = \sqrt{2(1 - \frac{17}{25})} = \frac{4}{5}$$

And for $|W\rangle = \frac{1}{\sqrt{3}}(|001\rangle + |010\rangle + |100\rangle)$:

$$\rho_M = \rho_{AB} = \rho_{AC} = \rho_{BC} = \begin{pmatrix} \frac{1}{3} & 0 & 0 & 0 \\ 0 & \frac{1}{3} & \frac{1}{3} & 0 \\ 0 & \frac{1}{3} & \frac{1}{3} & 0 \\ 0 & 0 & 0 & 0 \end{pmatrix}$$

$$C_M(\rho) = \sqrt{2(1 - Tr(\rho_M^2))} = \sqrt{2(1 - \frac{5}{9})} = \frac{2\sqrt{2}}{3}$$

We see that for all four compared states, the concurrence measure calculates more values than our measure.

## 4. Generalization to qudit states

Now we want to generalize the measure for 2-qudit states. The relation (8) for 2-qudit states is written as follows:

$$E = \lambda_{\max} - \gamma \qquad (9)$$

In this relation, the $\lambda_{\max}$ is the largest eigenvalue of the $\sigma$ operator for the desired dimension and it is used to adjust the value of the measure so that for separable states, the value $\gamma$ (separability index) is equal to $\lambda_{\max}$, to the measure value obtain zero. Also, for the fully entangled state, when the value of $\gamma$ becomes zero, the value of the measure will result in the largest eigenvalue. Because according to part 2 of this article, the separability index ($\gamma$) was defined in such a way that for the separable state, Alice and Bob found operators that would obtain the expectation value of the state equal to the maximum eigenvalue of the operator. It is obvious that the value of $\lambda_{\max}$ is 1 for qubit states, 2 for qutrit states, and $D-1$ for qudit states. D is the qudit dimension.

For N-qudit states, relation (2) should be corrected as follows:

$$\gamma = \frac{\alpha_1 |\max\langle\sigma_1\rangle - \eta_1| + \alpha_2 |\max\langle\sigma_2\rangle - \eta_2| + \ldots + \alpha_N |\max\langle\sigma_N\rangle - \eta_N|}{N} \qquad (10)$$

In fact, the target is to find the distance between the maximum expectation value of the generalized Pauli operator of the $i$nd component of the desired state ($\max\langle\sigma_i\rangle$) and the minimum absolute value of the expectation value that is specific to the fully entangled state $\eta_i$. Therefore, the value of $\eta_i$ is the average of the eigenvalues related to the states of the single particle (part of the state), which is calculated from the following relation:

$$\eta_i = \frac{1}{l}\left|\sum_j \lambda_{ij}\right| = \frac{|\lambda_{i1} + \lambda_{i2} + \ldots + \lambda_{il}|}{l} \qquad (11)$$

where $j$ is the number of state components that range from 1 to $l$ and $i$ is the number of desired qudit that varies from 1 to $N$. For example, in the 2-qutrit state $|\phi\rangle = \frac{1}{2}|02\rangle + \frac{\sqrt{3}}{4}|10\rangle + \frac{\sqrt{6}}{4}|20\rangle + \frac{\sqrt{2}}{4}|12\rangle$, the $i$ is between 1 and 2 (the number of qutrits) and the $j$ is between 1 and 4. The meaning of $\lambda_{ij}$ is the eigenvalue related to the $i$nd qutrit in the $j$nd component of the state. For example, in the latter state, the $\lambda_{23}$ is the eigenvalue related to the third component of the state (which is its coefficient is $\frac{\sqrt{6}}{4}$) and its second qutrit, which is $|0\rangle$. So $\lambda_{23} = 2$.

The parameter $\eta$ shows the expectation value of the operator in a fully entangled state made of the existing eigenstates. In fully entangled states, the expectation value is the average of the eigenvalues. For qubit states, since there are two eigenvalues 1 and -1, the $\eta$ becomes zero. But, in qutrit states, if the desired state is from the combination of $|0\rangle$ and $|1\rangle$ or $|1\rangle$ and $|2\rangle$, the value is equal to 1, and if it is from $|0\rangle$ and $|2\rangle$, this value becomes zero. If the state is a combination of all three $|0\rangle$ $|1\rangle$ and $|2\rangle$, the value of $\eta$ becomes zero.

This measure gives a value other than zero for some separable states. For example, the value of entanglement will not be zero for a state like $|\phi\rangle = \frac{1}{\sqrt{3}}|10\rangle + \frac{1}{\sqrt{3}}|11\rangle + \frac{1}{\sqrt{3}}|12\rangle$, but we know that this state is separable. Therefore, caution should be used in using this measure for separable states, and it is better to first by a method determine whether the state is entangled or separable and then use the measure for entangled states.

Finally, I must state the necessity of the $\alpha$ coefficient in equation (10). According to the fact that in relation (9), the $\lambda_{\max}$ is used in some states; For example, in $D=4$ (the qudit states with dimension 4), the eigenstate $|1\rangle$ where the eigenvalue is equal to 1 and it is different from the maximum eigenvalue that is 3, therefore, to calibrate the measure and to make the measure value equal to zero for the separable state, the $\alpha$ coefficient defined in the following relation is used:

$$\alpha_i = \frac{\lambda_{\max}}{\frac{1}{l}\sum_{j=1}^{l}|\lambda_{ij} - \eta_i|} = \frac{l\lambda_{\max}}{\sum_{j=1}^{l}|\lambda_{ij} - \eta_i|} \qquad (12)$$

In relation (10), the $\sigma_i$'s are the same generalized Pauli matrices that we write here their matrix representation for 3 and 4 dimensions in the computational basis to remind the reader :

$$\sigma.n = \begin{pmatrix} 2\cos\theta & \sqrt{2}e^{-i\varphi}\sin\theta & 0 \\ \sqrt{2}e^{i\varphi}\sin\theta & 0 & \sqrt{2}e^{-i\varphi}\sin\theta \\ 0 & \sqrt{2}e^{i\varphi}\sin\theta & -2\cos\theta \end{pmatrix} \qquad \text{for } D=3 \qquad (13)$$

$$\sigma.n = \begin{pmatrix} 3\cos\theta & \sqrt{3}e^{-i\varphi}\sin\theta & 0 & 0 \\ \sqrt{3}e^{i\varphi}\sin\theta & \cos\theta & 2e^{-i\varphi}\sin\theta & 0 \\ 0 & 2e^{i\varphi}\sin\theta & -\cos\theta & \sqrt{3}e^{-i\varphi}\sin\theta \\ 0 & 0 & \sqrt{3}e^{i\varphi}\sin\theta & -3\cos\theta \end{pmatrix} \qquad \text{for } D=4 \qquad (14)$$

***Example 3:*** The entanglement of 2- qutrit states:

Based on the relation (13), for $\sigma_1 = \sigma_1 \otimes I$ and $\sigma_2 = I \otimes \sigma_2$ we will have:

$$\sigma_1 = \begin{pmatrix} 2\cos\theta_1 & 0 & 0 & \sqrt{2}e^{-i\varphi_1}\sin\theta_1 & 0 & 0 & 0 & 0 & 0 \\ 0 & 2\cos\theta_1 & 0 & 0 & \sqrt{2}e^{-i\varphi_1}\sin\theta_1 & 0 & 0 & 0 & 0 \\ 0 & 0 & 2\cos\theta_1 & 0 & 0 & \sqrt{2}e^{-i\varphi_1}\sin\theta_1 & 0 & 0 & 0 \\ \sqrt{2}e^{i\varphi_1}\sin\theta_1 & 0 & 0 & 0 & 0 & 0 & \sqrt{2}e^{-i\varphi_1}\sin\theta_1 & 0 & 0 \\ 0 & \sqrt{2}e^{i\varphi_1}\sin\theta_1 & 0 & 0 & 0 & 0 & 0 & \sqrt{2}e^{-i\varphi_1}\sin\theta_1 & 0 \\ 0 & 0 & \sqrt{2}e^{i\varphi_1}\sin\theta_1 & 0 & 0 & 0 & 0 & 0 & \sqrt{2}e^{-i\varphi_1}\sin\theta_1 \\ 0 & 0 & 0 & \sqrt{2}e^{i\varphi_1}\sin\theta_1 & 0 & 0 & -2\cos\theta_1 & 0 & 0 \\ 0 & 0 & 0 & 0 & \sqrt{2}e^{i\varphi_1}\sin\theta_1 & 0 & 0 & -2\cos\theta_1 & 0 \\ 0 & 0 & 0 & 0 & 0 & \sqrt{2}e^{i\varphi_1}\sin\theta_1 & 0 & 0 & -2\cos\theta_1 \end{pmatrix}$$

$$\sigma_2 = \begin{pmatrix} 2\cos\theta_2 & \sqrt{2}e^{-i\varphi_2}\sin\theta_2 & 0 & 0 & 0 & 0 & 0 & 0 & 0 \\ \sqrt{2}e^{i\varphi_2}\sin\theta_2 & 0 & \sqrt{2}e^{-i\varphi_2}\sin\theta_2 & 0 & 0 & 0 & 0 & 0 & 0 \\ 0 & \sqrt{2}e^{i\varphi_2}\sin\theta_2 & -2\cos\theta_2 & 0 & 0 & 0 & 0 & 0 & 0 \\ 0 & 0 & 0 & 2\cos\theta_2 & \sqrt{2}e^{-i\varphi_2}\sin\theta_2 & 0 & 0 & 0 & 0 \\ 0 & 0 & 0 & \sqrt{2}e^{i\varphi_2}\sin\theta_2 & 0 & \sqrt{2}e^{-i\varphi_2}\sin\theta_2 & 0 & 0 & 0 \\ 0 & 0 & 0 & 0 & \sqrt{2}e^{i\varphi_2}\sin\theta_2 & -2\cos\theta_2 & 0 & 0 & 0 \\ 0 & 0 & 0 & 0 & 0 & 0 & 2\cos\theta_2 & \sqrt{2}e^{-i\varphi_2}\sin\theta_2 & 0 \\ 0 & 0 & 0 & 0 & 0 & 0 & \sqrt{2}e^{i\varphi_2}\sin\theta_2 & 0 & \sqrt{2}e^{-i\varphi_2}\sin\theta_2 \\ 0 & 0 & 0 & 0 & 0 & 0 & 0 & \sqrt{2}e^{i\varphi_2}\sin\theta_2 & -2\cos\theta_2 \end{pmatrix}$$

If we want to calculate entanglement for state $|\phi\rangle = \frac{1}{2}|02\rangle + \frac{\sqrt{3}}{2}|20\rangle$, we have:

$$\langle\sigma_1\rangle = -\cos\theta_1 \quad \langle\sigma_2\rangle = \cos\theta_2 \quad , \quad \eta_1 = \frac{|\lambda_0 + \lambda_2|}{2} = \frac{|2+-2|}{2} = 0 \quad \eta_2 = \frac{|\lambda_2 + \lambda_1|}{2} = \frac{|-2+2|}{2} = 0$$

$$\alpha_1 = \frac{l\lambda_{max}}{\sum_{j=1}^{l}|\lambda_{1j} - \eta_1|} = \frac{2*2}{|2-0| + |-2-0|} = 1 \quad \alpha_2 = \frac{l\lambda_{max}}{\sum_{j=1}^{l}|\lambda_{2j} - \eta_2|} = \frac{2*2}{|2-0| + |-2-0|} = 1$$

$$E = \lambda_{max} - \frac{\alpha_1 |\max\langle\sigma_1\rangle - \eta_1| + |\max\langle\sigma_2\rangle - \eta_2|}{2} = 2 - \frac{1+1}{2} = 1$$

And for $|\phi\rangle = \frac{1}{2}|01\rangle + \frac{\sqrt{3}}{2}|20\rangle$:

$$\langle\sigma_1\rangle = -\cos\theta_1 \qquad \langle\sigma_2\rangle = \frac{3}{2}\cos\theta_2 \quad , \quad \eta_1 = \frac{|\lambda_0 + \lambda_2|}{2} = \frac{|2 + -2|}{2} = 0 \qquad \eta_2 = \frac{|\lambda_1 + \lambda_0|}{2} = \frac{|0 + 2|}{2} = 1$$

$$\alpha_1 = \frac{l\lambda_{max}}{\sum_{j=1}^{l}|\lambda_{1j} - \eta_1|} = \frac{2*2}{|2-0| + |-2-0|} = 1 \qquad \alpha_2 = \frac{l\lambda_{max}}{\sum_{j=1}^{l}|\lambda_{2j} - \eta_2|} = \frac{2*2}{|0-1| + |2-1|} = 2$$

$$E = \lambda_{max} - \frac{\alpha_1|\max\langle\sigma_1\rangle - \eta_1| + |\max\langle\sigma_2\rangle - \eta_2|}{2} = 2 - \frac{1+1}{2} = 1$$

At a glance, we can see that the above two states have equal entanglement. During the calculation, we obtained different numbers for the expectation values and $\eta$ and $\alpha$, but finally, we saw that the entanglement values were equal. Now we will check some other states:

***Example 4:*** Now, for the two-qutrit state which is $l = 3$, for example, the state $|\phi\rangle = \frac{1}{\sqrt{3}}|00\rangle + \frac{1}{\sqrt{3}}|11\rangle + \frac{1}{\sqrt{3}}|22\rangle$:

$$\langle\sigma_1\rangle = 0 \qquad \langle\sigma_2\rangle = 0 \quad , \quad \eta_1 = \frac{|\lambda_0 + \lambda_1 + \lambda_2|}{3} = \frac{|2 + 0 - 2|}{3} = 0 \qquad \eta_2 = \frac{|\lambda_0 + \lambda_1 + \lambda_2|}{3} = \frac{|2 + 0 - 2|}{3} = 0$$

$$\alpha_1 = \frac{l\lambda_{max}}{\sum_{j=1}^{l}|\lambda_{1j} - \eta_1|} = \frac{3*2}{|2| + |0| + |-2|} = \frac{3}{2} \qquad \alpha_2 = \frac{l\lambda_{max}}{\sum_{j=1}^{l}|\lambda_{2j} - \eta_2|} = \frac{3*2}{|2| + |0| + |-2|} = \frac{3}{2}$$

$$E = \lambda_{max} - \frac{\alpha_1|\max\langle\sigma_1\rangle - \eta_1| + \alpha_2|\max\langle\sigma_2\rangle - \eta_2|}{2} = 2 - \frac{\frac{3}{2}|0-0| + \frac{3}{2}|0-0|}{2} = 2$$

In this example, we know that due to the fact that the expectation values and $\eta$ have become zero, the $\alpha$ values are ineffective, so we can not calculate them.

Taking into account that for calculation $\eta$, we obtain the average of the possible eigenvalues for the desired qudit, therefore, we note that we avoid using $\lambda$ repetitious in $\eta$ calculations, otherwise, there will be an error in the calculations. For example in state $|\phi\rangle = \frac{1}{\sqrt{3}}|00\rangle + \frac{1}{\sqrt{2}}|11\rangle + \frac{1}{\sqrt{6}}|20\rangle$, this is how we act:

$$\langle \sigma_1 \rangle = \frac{1}{3}\cos\theta_1 \qquad \langle \sigma_2 \rangle = \cos\theta_2 \quad , \quad \eta_1 = \frac{|\lambda_0 + \lambda_1 + \lambda_2|}{3} = \frac{|2+0-2|}{3} = 0 \qquad \eta_2 = \frac{|\lambda_0 + \lambda_1|}{2} = \frac{|2+0|}{2} = 1$$

$$\alpha_1 = \frac{l\lambda_{\max}}{\sum_{j=1}^{l}|\lambda_{1j} - \eta_1|} = \frac{3*2}{|2|+|0|+|-2|} = \frac{3}{2} \qquad \alpha_2 = \frac{l\lambda_{\max}}{\sum_{j=1}^{l}|\lambda_{2j} - \eta_2|} = \frac{3*2}{|2-1|+|0-1|+|2-1|} = 2$$

$$E = \lambda_{\max} - \frac{\alpha_1|\max\langle\sigma_1\rangle - \eta_1| + \alpha_2|\max\langle\sigma_2\rangle - \eta_2|}{2} = 2 - \frac{\frac{3}{2}\left|\frac{1}{3} - 0\right| + 2|1-1|}{2} = \frac{7}{4}$$

***Example 5:*** The entanglement of qudit states of $D = 4$:

For the 2-qudit state $|\phi\rangle = \frac{1}{\sqrt{5}}|01\rangle + \frac{2}{\sqrt{5}}|10\rangle$, we calculate:

$$\langle \sigma_1 \rangle = \frac{7}{5}\cos\theta_1 \qquad \langle \sigma_2 \rangle = \frac{13}{5}\cos\theta_2 \quad , \quad \eta_1 = \frac{|\lambda_0 + \lambda_1|}{2} = \frac{|3+1|}{2} = 2 \qquad \eta_2 = \frac{|\lambda_1 + \lambda_0|}{2} = \frac{|1+3|}{2} = 2$$

$$\alpha_1 = \frac{l\lambda_{\max}}{\sum_{j=1}^{l}|\lambda_{1j} - \eta_1|} = \frac{2*3}{|1-2|+|3-2|} = 3 \qquad \alpha_2 = \frac{l\lambda_{\max}}{\sum_{j=1}^{l}|\lambda_{2j} - \eta_2|} = \frac{2*3}{|1-2|+|3-2|} = 3$$

$$E = \lambda_{\max} - \frac{\alpha_1|\max\langle\sigma_1\rangle - \eta_1| + \alpha_2|\max\langle\sigma_2\rangle - \eta_2|}{2} = 3 - \frac{3\left|\frac{7}{5} - 2\right| + 3\left|\frac{13}{5} - 2\right|}{2} = \frac{6}{5}$$

And for the 2-qudit state $|\phi\rangle = \frac{1}{\sqrt{5}}|03\rangle + \frac{2}{\sqrt{5}}|30\rangle$:

$$\langle \sigma_1 \rangle = -\frac{9}{5}\cos\theta_1 \quad , \quad \langle \sigma_2 \rangle = \frac{9}{5}\cos\theta_2 \quad , \quad \eta_1 = \frac{|\lambda_0 + \lambda_3|}{2} = \frac{|3+-3|}{2} = 0 \quad , \quad \eta_2 = \frac{|\lambda_3 + \lambda_0|}{2} = \frac{|-3+3|}{2} = 0$$

$$\alpha_1 = \frac{l\lambda_{max}}{\sum_{j=1}^{l}|\lambda_{1j}-\eta_1|} = \frac{2*3}{|3|+|-3|} = 1 \qquad \alpha_2 = \frac{l\lambda_{max}}{\sum_{j=1}^{l}|\lambda_{2j}-\eta_2|} = \frac{2*3}{|-3|+|3|} = 1$$

$$E = \lambda_{max} - \frac{\alpha_1|\max\langle\sigma_1\rangle-\eta_1|+\alpha_2|\max\langle\sigma_2\rangle-\eta_2|}{2} = 3 - \frac{1\left|\frac{9}{5}-0\right|+1\left|\frac{9}{5}-0\right|}{2} = \frac{6}{5}$$

Also for the 2-qudit state $|\phi\rangle = \frac{1}{\sqrt{5}}|12\rangle + \frac{2}{\sqrt{5}}|21\rangle$:

$$\langle\sigma_1\rangle = -\frac{3}{5}\cos\theta_1 \quad , \quad \langle\sigma_2\rangle = \frac{3}{5}\cos\theta_2 \quad , \quad \eta_1 = \frac{|\lambda_1+\lambda_2|}{2} = \frac{|1+-1|}{2} = 0 \quad , \quad \eta_2 = \frac{|\lambda_2+\lambda_1|}{2} = \frac{|-1+1|}{2} = 0$$

$$\alpha_1 = \frac{l\lambda_{max}}{\sum_{j=1}^{l}|\lambda_{1j}-\eta_1|} = \frac{2*3}{|1-0|+|-1-0|} = 3 \qquad \alpha_2 = \frac{l\lambda_{max}}{\sum_{j=1}^{l}|\lambda_{2j}-\eta_2|} = \frac{2*3}{|-1-0|+|1-0|} = 3$$

$$E = \lambda_{max} - \frac{\alpha_1|\max\langle\sigma_1\rangle-\eta_1|+\alpha_2|\max\langle\sigma_2\rangle-\eta_2|}{2} = 3 - \frac{3\left|\frac{3}{5}-0\right|+3\left|\frac{3}{5}-0\right|}{2} = \frac{6}{5}$$

In example 5, we can see that despite the different values of the parameters $\eta$ and $\alpha$ as well as the expectation values for different states, the same values of entanglement have been obtained, which is expected and this shows that the relationships correctly define an efficient measure.

## 5. Conclusion:

The introduced measure is based on the concept of entanglement and its effect on the expectation values of Pauli operators. According to the mentioned examples, this measure determines the amount of entanglement of N-qubit states well. Also, this measure calculates the entanglement of qudit states well. The measure value is equal to $\lambda_{max}$ for fully entangled states, which is equal to 1 for qubit states and equal to 2 for qutrit states, and in general, for other qudit states, it is equal to the $D-1$ value, which $D$ is the qudit dimension.